\documentclass[11pt,onecolumn,a4paper]{article}

\usepackage[margin=1in]{geometry}
\usepackage{cite}
\usepackage{amsmath,amssymb,amsfonts}
\usepackage{algorithmic}
\usepackage{graphicx}
\usepackage{textcomp}
\usepackage{wrapfig}
\usepackage{booktabs}
\usepackage{pifont}
\usepackage{xcolor}
\usepackage{multirow}
\usepackage{threeparttable}
\usepackage{enumitem}
\usepackage[hyphens]{url}
\usepackage[hidelinks]{hyperref}
\usepackage{authblk}

\begin{document}

\title{Synergistic Blood Pressure Estimation via Contactless mmWave Radar and Imaging Photoplethysmography: A Feasibility Study}

\author[1]{Boyuan Gu}
\author[2]{Haiyang Sun\thanks{Equal contribution.}}
\author[2]{Yongjie Liu\thanks{Equal contribution.}}
\author[2]{Guangji Ma}
\author[3]{Guanghao Sun}
\author[2,4]{Xiaorong Ding\thanks{Corresponding author: xiaorong.ding@uestc.edu.cn}}

\affil[1]{Glasgow College, University of Electronic Science and Technology of China, Chengdu, China}
\affil[2]{School of Life Science and Technology, University of Electronic Science and Technology of China, Chengdu, China}
\affil[3]{Graduate School of Informatics and Engineering, The University of Electro-Communications, Chofu, Japan}
\affil[4]{Yangtze Delta Region Institute (Huzhou), University of Electronic Science and Technology of China, Huzhou, China}

\maketitle

\begin{abstract}
Continuous, non-contact blood pressure (NCBP) monitoring holds significant promise for pervasive cardiovascular care, yet single-modality approaches---such as imaging photoplethysmography (iPPG)---remain constrained by environmental artifacts, skin-tone sensitivity, and the absence of proximal cardiac mechanical information. This study investigates the feasibility of a dual-modality sensing paradigm that synergistically integrates facial iPPG with posterior-facing frequency-modulated continuous wave (FMCW) millimeter-wave radar to capture complementary hemodynamic cues: distal optical volumetric fluctuations and proximal cardiac micro-motions (radar motion signals, RMS). To bridge the morphological disparity between these heterogeneous streams, we develop an end-to-end deep learning architecture, BiLSTM-MS-DiCNN, which leverages multi-scale dilated convolutions for spatial feature extraction and bidirectional long short-term memory for temporal dependency modeling. In a controlled feasibility study involving 15 healthy participants across distinct hemodynamic states (resting, deep breathing, and post-exercise), the proposed framework achieved a Mean Absolute Difference (MAD) of 4.71 mmHg for systolic BP (SBP) and 4.60 mmHg for diastolic BP (DBP) under resting conditions, with consistent performance during physiological perturbations (SBP: 6.35 mmHg, DBP: 4.95 mmHg for deep breathing; SBP: 5.33 mmHg, DBP: 4.96 mmHg for post-exercise). Statistically significant improvements over unimodal baselines validate the complementary nature of the dual-modality inputs, while ablation studies confirm the architectural contributions of each module. Even without calibration (0\%), the model achieves SBP/DBP MAD of 7.52/6.85 mmHg, indicating that the framework retains a measurable non-calibrated baseline capability, while brief calibration further improves subject-level personalization. These preliminary findings demonstrate the viability of mmWave-iPPG fusion as a promising pathway toward robust, unobtrusive NCBP monitoring.

\end{abstract}

\noindent\textit{Keywords}: Non-contact Blood Pressure Measurement, Imaging Photoplethysmography, mmWave Radar, Dual-Modality, BiLSTM, DiCNN
```latex
\section{Introduction}
\label{sec:introduction}
Cardiovascular diseases, commonly resulting from high blood pressure (BP) or hypertension, remain a predominant global health challenge, accounting for a substantial proportion of non-communicable disease mortality \cite{1murray2020global}. Often termed the ``silent killer,'' hypertension is characterized by a high rate of undiagnosed cases, with the World Health Organization (WHO) reporting an awareness rate of 46\%, indicating that more than half of at-risk individuals tend to ignore their hypertension conditions.  \cite{2gnanenthiran2024shop,3yang2018prevalence, 4cao2022burden}. This alarming figure reflects a worrying fact of underestimated hypertension risk. Frequent and timely BP measurement is crucial for the accurate prediction and management of hypertension. Traditional BP measurement methods can be broadly categorized into invasive and non-invasive techniques. Invasive BP measurement involves insertions into the artery to capture hemodynamic-based signals, providing precise measurement results. Although this method is indispensable, its application is limited to clinical uses due to its invasive nature \cite{6peters2024invasively}. Non-invasive techniques, typically using an arterial cuff, offer safer and more convenient alternatives. Cuff-based devices, though portable, are restricted to ``snapshot'' readings and often induce physical discomfort, leading to poor user compliance.  In recent years, cuff-less BP estimation has emerged as a promising approach within non-invasive methods, leveraging physiological signals such as the electrocardiogram (ECG) and photoplethysmogram (PPG) for BP prediction \cite{gu2025unettransformer}. Both ECG and PPG signals are widely used in BP estimation models by deriving crucial features including pulse arrival time (PAT) \cite{0908-1} and pulse transit time (PTT) \cite{0908-2, gu2026viscoelastic}, particularly for the prediction of SBP and DBP. However, the persistent reliance on contact electrodes or finger clamps imposes a ``wearable burden'' that precludes continuous, free-living monitoring \cite{13contactless2024blood, 14maurya2021non}. Furthermore, these methods primarily capture vascular dynamics rather than cardiac mechanical signals, leading to drift and limited accuracy during prolonged measurement. Therefore, there is a critical clinical and practical need for NCBP monitoring techniques that can overcome these fundamental limitations.

Despite independent advancements, single-modality NCBP systems encounter inherent physical bottlenecks that compromise their reliability in unconstrained environments. For instance, classical imaging photoplethysmography (iPPG) captures peripheral blood volume changes by analyzing regions of interest (ROI) in facial videos \cite{0908-3, 0908-4, 0908-5}. While deep learning architectures, such as U-Net, have successfully transformed iPPG into continuous BP waveforms with clinically acceptable errors (e.g., a MAD of 7.52 mmHg for MAP) \cite{16bousefsaf2022estimation}, this optical modality is heavily constrained by environmental factors. Its signal fidelity is severely degraded by unstable ambient illumination and variations in skin melanin concentration, posing significant challenges to algorithmic robustness across diverse populations \cite{15park2024robust, 17rong2021blood, 0908-6}. Conversely, frequency-modulated continuous wave (FMCW) mmWave radar demonstrates high precision in capturing sub-millimeter chest-wall micromotions---yielding the radar motion signal (RMS) that reflects cardiac mechanics \cite{18murshed2023cnn}. Although mmWave radar alone has achieved promising NCBP estimations with a MAD of approximately 5 mmHg \cite{0908-7}, it fundamentally lacks the peripheral arterial network visibility required to compute transit-based BP surrogates independently, often necessitating complex distributed measurement setups \cite{0908-8}. 

Consequently, a synergistic dual-modality paradigm is fundamentally requisite. Biomechanically, FMCW radar and iPPG provide complementary boundary conditions of the cardiovascular tree. The radar identifies the proximal cardiac ejection event \cite{wang2020sensors, lv2021sensors}, while iPPG records the distal pulse arrival \cite{0908-3}. Time-aligning these modalities yields a non-contact Pulse Transit Time (ncPTT) surrogate, a physiological parameter inversely correlated with blood pressure \cite{mukkamala2015tbme}. Recent exploratory studies have validated the feasibility of integrating radar and PPG for cuff-less BP monitoring \cite{cho2021thc, zhang2023fusion, ma2025multimodal}. However, a critical gap remains in the methodological fusion of these heterogeneous signals. Existing dual-modality frameworks predominantly rely on rudimentary feature-level integration---such as computing a scalar PTT via traditional peak-detection algorithms, or utilizing shallow machine learning models for post-hoc feature concatenation. This shallow fusion is highly susceptible to transient motion artifacts and signal phase misalignment. Furthermore, the profound morphological disparity between one-dimensional radar micro-motions and high-dimensional iPPG optical fluctuations presents a significant spatio-temporal challenge. The current literature lacks a systematic feasibility investigation that employs a sophisticated deep-learning paradigm capable of concurrently extracting cross-modal spatial representations and decoding the complex, long-range temporal dependencies hidden within these dual physiological streams.

To address this gap, this study investigates the feasibility of a dual-modality NCBP estimation framework that synergistically integrates facial iPPG and posterior-facing mmWave radar RMS. By simultaneously monitoring the face and back, our sensing configuration isolates distal optical volumetric fluctuations from proximal mechanical cardiac vibrations, minimizing spatial cross-interference. At the core of our framework is a deep-learning network, BiLSTM-MS-DiCNN, designed to implicitly learn the non-linear hemodynamic mapping from heterogeneous physiological streams. Instead of relying on fragile, intermediate peak-detection algorithms to extract a scalar PTT, our end-to-end approach directly maps the synchronized dual-modality inputs to continuous SBP and DBP values. The key contributions of this feasibility study are summarized as follows:

\begin{enumerate} \item \textbf{Exploratory Dual-Modality Sensing Paradigm:} We systematically investigate a face-back distributed NCBP measurement approach combining mmWave radar and iPPG, examining whether this physical configuration can effectively mitigate single-modality environmental sensitivity while providing complementary boundary conditions of the cardiovascular tree. \item \textbf{Deep Learning Architecture for Heterogeneous Fusion:} We develop the hybrid BiLSTM-MS-DiCNN network to address the morphological disparity between optical and radar signals, and validate its effectiveness through comprehensive ablation studies. \item \textbf{Feasibility Assessment Across Hemodynamic States:} We evaluate the framework on a curated dataset involving 15 healthy participants across distinct cardiovascular states---resting, deep breathing, and post-exercise---demonstrating the potential of dual-modality fusion for tracking dynamic physiological fluctuations. \item \textbf{Quantitative Cross-Modal Synergy Analysis:} Through extensive comparative experiments against single-modality baselines and mainstream models, we provide quantitative evidence of the complementary nature of iPPG and mmWave radar inputs for NCBP estimation. \end{enumerate}
```

\begin{figure}[t]
    \centering
    \includegraphics[width=1\linewidth]{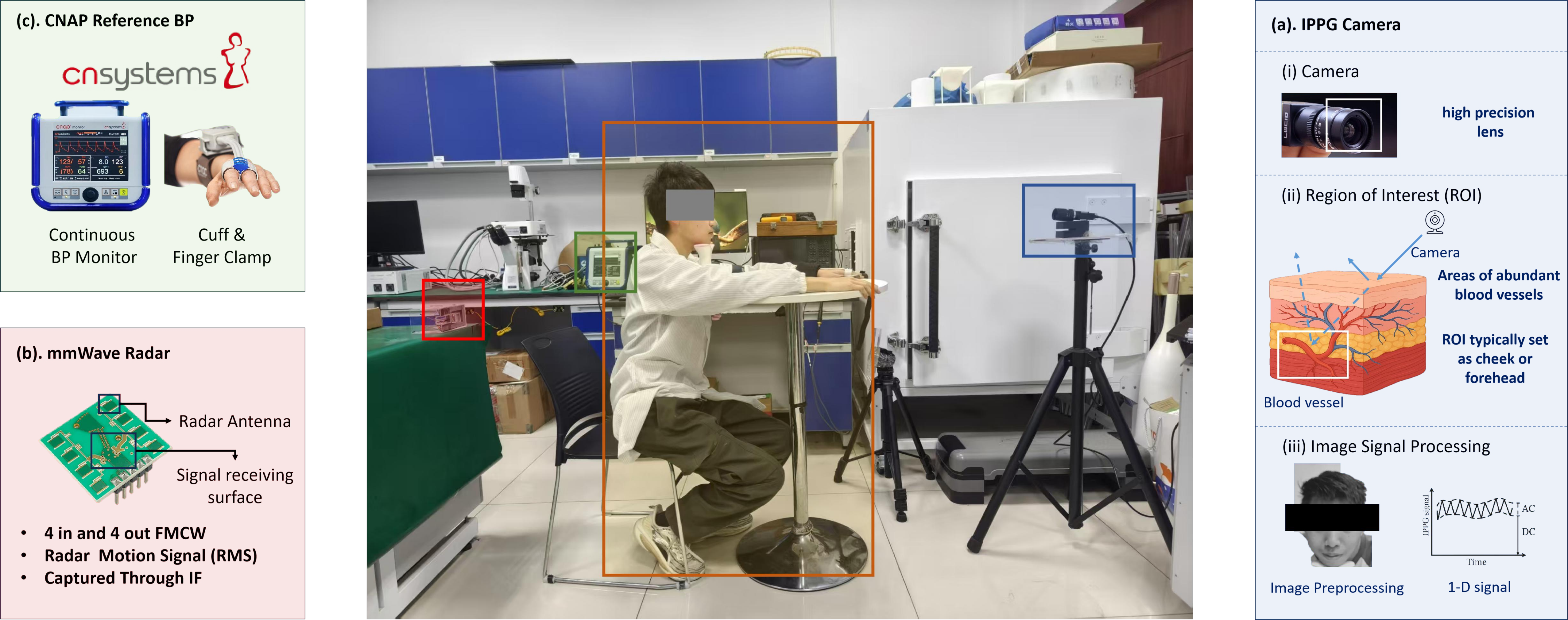}
    \caption{Laboratory setup for signal acquisition: (a). high precision camera for iPPG, (b) mmWave radar for RMS, (c) CNAP for reference BP.}
    \label{fig1:expdesign}
\end{figure}

\section{Methodology}

To achieve robust, continuous NCBP estimation, our methodology is constructed upon a physiologically-driven pipeline. As illustrated in the experimental setup (Figure 1), we strategically established a dual-modality signal capturing pipeline via iPPG and mmWave radar. Section II.A first models the mathematical foundation of this dual-modality fusion, linking ncPTT to blood pressure variations utilizing the Moens-Korteweg (M-K) principle. Based on this physiological modeling, we delineate the signal processing mechanisms employed to reliably extract and synchronize proximal cardiac micro-motions (via mmWave radar) and distal blood volume changes (via facial iPPG) in Section II.B. Following this, we introduce the uniquely architected BiLSTM-MS-DiCNN (Section II.C). This end-to-end network is explicitly designed to overcome cross-modal morphological disparities, translating the raw heterogeneous physiological streams into accurate SBP and DBP estimations without relying on fragile heuristic feature engineering.

\subsection{Physiological Basis of Dual-Modality Fusion}

The theoretical foundation of proposed NCBP estimation is governed by the biomechanical relationship between BP and Pulse Wave Velocity (PWV), which is classically modeled by the combination of the M-K and Hughes equations. Given that PWV is determined by the arterial propagation distance ($L$) divided by the Pulse Transit Time (PTT), the non-linear relationship between BP and PTT can be mathematically approximated as:

\begin{equation}
    BP = \frac{a}{PTT^2} + b
\end{equation}

where $a$ and $b$ are subject-specific parameters related to arterial compliance, vessel elasticity, and blood density. 

This physiological mapping dictates that extracting an accurate, robust PTT surrogate is the fundamental prerequisite for continuous BP monitoring. Physiologically, PTT is defined as the temporal interval between the proximal ejection of blood from the heart---specifically initiated by the aortic valve opening (AVO)---and the arrival of the consequent pulse wave at a distal peripheral site. To construct a complete, ncPTT measurement loop without physical tethering, our framework leverages two complementary sensing principles to capture these physiological boundary conditions. First, the FMCW mmWave radar acquires the sub-millimeter mechanical micro-motions of the torso. This RMS reflects the proximal cardiomechanical vibrations, capturing the initial systolic ejection event. Concurrently, facial iPPG captures the optical absorption variations induced by the rhythmic blood volume changes within the dense capillary bed of the face, effectively registering the distal pulse arrival.

While the mathematical correlation between PTT and BP is explicit, translating this theoretical model into an unconstrained non-contact scenario presents profound methodological challenges. Traditional dual-modality approaches attempt to explicitly calculate a scalar PTT by detecting heuristic fiducial points (e.g., peaks or valleys) across the heterogeneous iPPG and RMS waveforms. However, this shallow feature-level fusion is highly susceptible to phase misalignments, transient motion artifacts, and the fundamental morphological disparity between one-dimensional mechanical radar waves and complex optical fluctuations. To overcome this, our methodology fundamentally shifts from explicit scalar extraction to implicit spatio-temporal mapping. By utilizing a uniquely architected deep learning network (detailed in Section II.C), the proposed system directly learns the non-linear hemodynamic dynamics governed by the M-K principle. The network adaptively aligns the multi-scale morphological representations of the proximal RMS and distal iPPG, mapping the raw, synchronized dual-modality streams directly to continuous SBP and DBP values.

\subsection{Dual-Modality Signal Extraction and Synchronization}

\begin{figure}[h]
    \centering
    \includegraphics[width=1\linewidth]{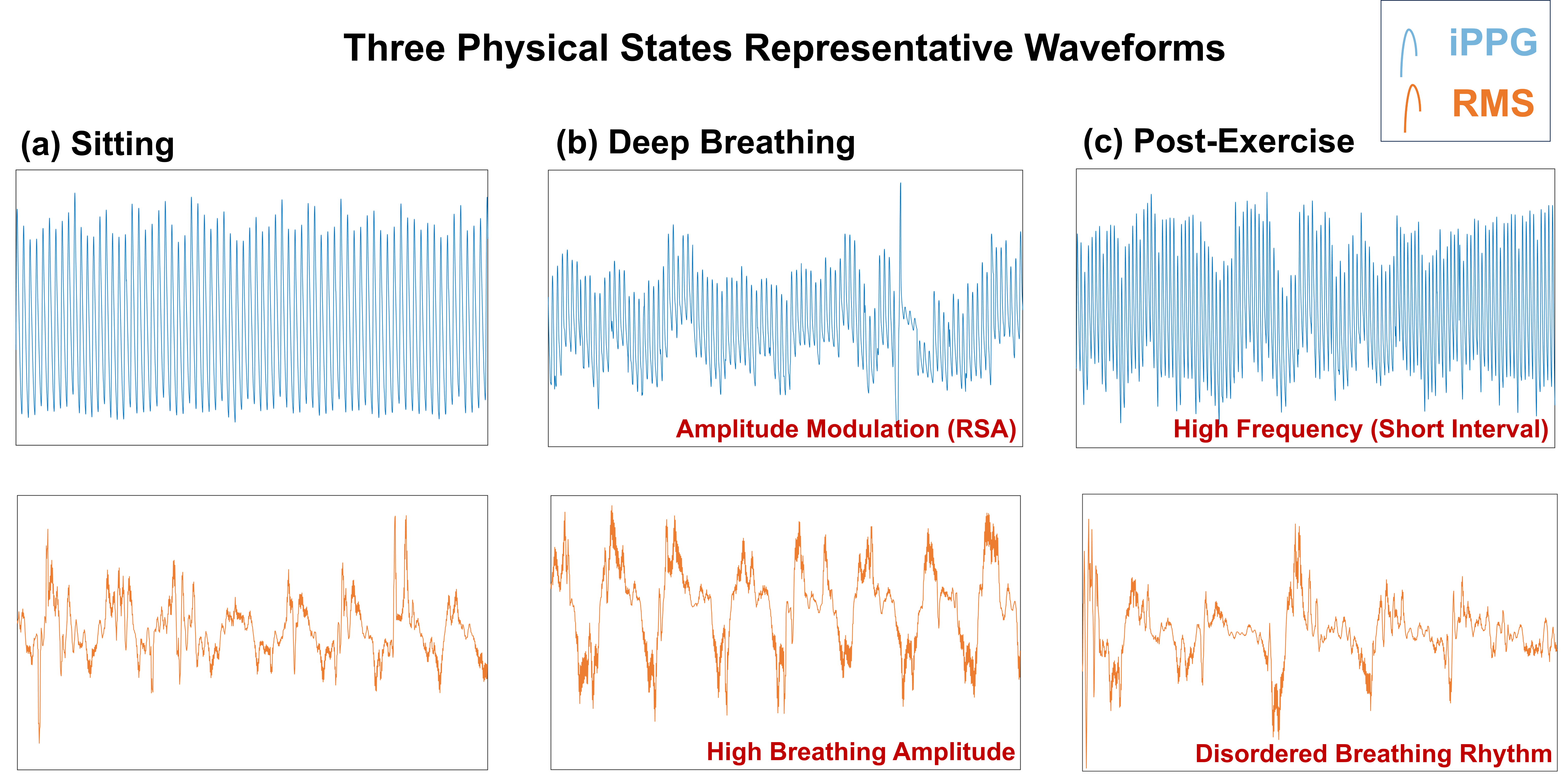}
    \caption{Representative waveforms of iPPG (blue) and RMS (orange) signals across three physiological states. (a) Sitting: Stable baseline with regular cardiac and respiratory rhythms. (b) Deep Breathing: Characterized by significant Respiratory Sinus Arrhythmia (RSA) amplitude modulation in iPPG and high breathing amplitude in RMS. (c) Post-Exercise: Exhibits high-frequency cardiac peaks (short intervals) and disordered breathing rhythms during recovery.}
    \label{fig:ippgandrms}
    \vspace{-1em}
\end{figure}

To operationalize the physiological framework established in Section II.A, robust extraction of the proximal cardiac micro-motions and distal optical volume changes is required. Each raw sensor channel undergoes tailored noise reduction to isolate the relevant hemodynamic features for subsequent spatio-temporal fusion. The representative waveforms of the processed imaging photoplethysmography (iPPG) and Radar Motion Signal (RMS) across diverse physiological states are illustrated in Figure \ref{fig:ippgandrms}.

\subsubsection{Distal iPPG Signal Extraction}
The iPPG signal, representing the distal pulse arrival, was extracted from the facial Region of Interest (ROI) centered on the forehead. Following ROI localization, the Red--Green--Blue (RGB) channels were projected using the CHROM combination \cite{1010-1} to yield a one-dimensional (1D) optical trace. This trace was zero-phase filtered to prevent phase distortion, which is critical for preserving pulse timing. To further enhance signal fidelity, Successive Variational Mode Decomposition (SVMD) was applied to the 1D iPPG signal with \textit{maxAlpha} set to 20,000, an initial center frequency of 0, and a stopping criterion where the power difference between the consecutive extracted modes fell below 0.0002. This decomposition effectively mitigates illumination artifacts and stabilizes the volumetric baseline.

\subsubsection{Proximal RMS Extraction}
The proximal cardiac-related mechanical motion was derived from the complex baseband I/Q stream of the mmWave radar. Let $I(t)$ and $Q(t)$ denote the in-phase and quadrature components; the complex signal is defined as:
\begin{equation}
s(t)=I(t)+jQ(t)=A(t)\,e^{j\phi(t)} ,
\end{equation}
from which the instantaneous (wrapped) phase is computed as:
\begin{equation}
\phi_{\mathrm{wrap}}(t)=\operatorname{atan2}\!\big(Q(t),\,I(t)\big),
\end{equation}
and subsequently unwrapped to obtain a continuous phase trajectory:
\begin{equation}
\phi(t)=\operatorname{unwrap}\!\big(\phi_{\mathrm{wrap}}(t)\big).
\end{equation}
For the FMCW radar, the sub-millimeter chest-wall displacement is strictly proportional to this phase, formulated as $x(t)=\tfrac{\lambda}{4\pi}\phi(t)$, where $\lambda$ is the carrier wavelength.

The unwrapped phase was zero-meaned and filtered utilizing a \emph{zero-phase}, 4th-order Butterworth low-pass filter with a 4~Hz cutoff to suppress high-frequency environmental clutter. Spectral energy analysis confirms that this cutoff preserves 99.9\% of the cardiac band energy (0.8--2~Hz) while strongly attenuating noise above 4~Hz, verifying that no physiologically relevant high-frequency morphological information is lost. Subsequently, SVMD was applied (\textit{maxAlpha} = 20,000, relative energy change stopping criterion $< 2\times10^{-4}$). To explicitly reconstruct the cardiomechanical component representing the initial systolic ejection, intrinsic modes with center frequencies within \mbox{0.8--2~Hz} (corresponding to an expected dynamic heart rate of $\sim$48--120~bpm) were isolated and recombined. The \textit{maxAlpha} parameter controls the fidelity of mode decomposition; a value of 20,000 was selected via grid search over the range $[1{,}000,\;50{,}000]$ on a pilot subset of three subjects, optimizing for the spectral separation between cardiac and respiratory components. This setting was subsequently applied uniformly across all subjects and physiological states without requiring state-specific tuning, as the fundamental frequency separation between cardiac (0.8--2~Hz) and respiratory (0.1--0.5~Hz) components remains stable regardless of the hemodynamic perturbation.

\subsubsection{Ground Truth Calibration and Inter-Modal Synchronization}
The reference continuous BP waveforms were recorded concurrently utilizing the clinical-grade CNAP system. The upper and lower envelope lines of the calibrated CNAP signals were computed to extract the beat-to-beat Systolic Blood Pressure (SBP) and Diastolic Blood Pressure (DBP) values, serving as the ground truth for model supervision.

To ensure strict temporal alignment for ncPTT mapping, both the extracted iPPG and RMS streams were uniformly downsampled to a sampling frequency of 30 Hz. To mitigate potential edge effects introduced during filtering and downsampling, a 5-second buffer was truncated from both extremities of the signals, yielding a stable, continuous 290-second duration for each modality. We opted for truncation over symmetric or reflection padding because the post-exercise protocol captures rapid hemodynamic recovery dynamics at the recording boundaries; padding techniques would introduce synthetic signal content that could be misinterpreted by the network as physiological activity, whereas truncation preserves data integrity at the cost of a modest reduction in usable duration. This rigorous synchronization guarantees that the heterogenous inputs fed into the deep learning architecture share an identical temporal receptive field, which is indispensable for the network to capture phase differences precisely.

\subsection{Physiologically-Driven Deep Learning Architecture}

\begin{figure}[t]
    \centering
    \includegraphics[width=1\linewidth]{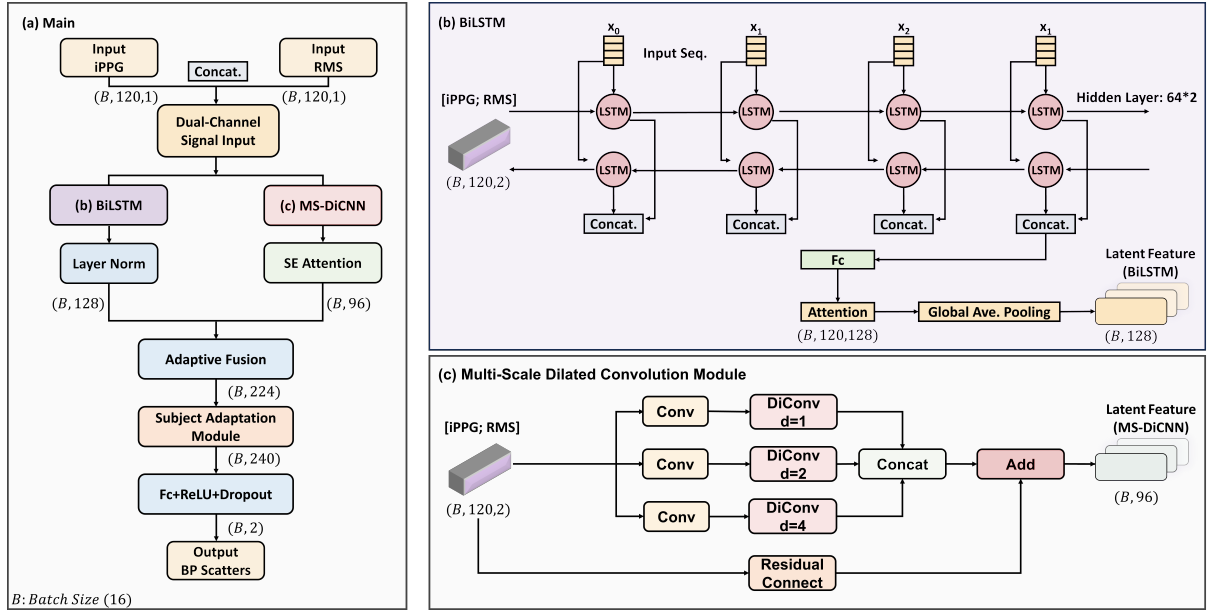}
    \caption{The proposed deep learning framework for continuous BP estimation. (a) Overview: Illustrates the parallel extraction and fusion of temporal and morphological features, the Subject Adaptation Module (SAM) to mitigate inter-subject variability. (b) BiLSTM Module: Extracts long-term temporal contexts using bidirectional states and global pooling. (c) MS-DiCNN Module: Captures multi-scale structural details via dilated convolutions.}
    \label{fig3:architecture}
\end{figure}

As established in Section II.A, continuous BP estimation relies on mapping the ncPTT surrogate to dynamic BP variations. However, explicitly extracting scalar time intervals from heterogenous signals is highly susceptible to morphological disparities and environmental artifacts. To implicitly solve this non-linear mapping end-to-end, we propose the BiLSTM-MS-DiCNN architecture (illustrated in Figure \ref{fig3:architecture}). 

The network takes the synchronized distal iPPG and proximal RMS signals as dual-channel inputs. The sequences are windowed into 120 time steps (4 seconds at 30 Hz), a duration strategically selected to encompass multiple complete cardiac cycles, thereby capturing both high-frequency systolic peaks and low-frequency RSA. The input tensor of size $(B, 120, 2)$ is processed in parallel by two specialized branches designed to decouple the spatial and temporal complexities of the cardiovascular system. 

\subsubsection{Temporal Dependency Modeling via BiLSTM}
Blood pressure regulation is not a memoryless process; it is continuously modulated by the autonomic nervous system (ANS) and is subject to vascular wave reflections from previous cardiac cycles. To mirror this physiological memory, the BiLSTM branch is employed to decode the long-range hemodynamic temporal dependencies. By processing the physiological sequence in both forward and backward directions, the BiLSTM effectively contextualizes the transient cardiac events within the broader regulatory rhythm. A Global Average Pooling layer subsequently aggregates this temporal evolution into a robust, high-level latent representation.

\subsubsection{Multi-Scale Morphological Alignment via MS-DiCNN}
The primary challenge of dual-modality fusion lies in the profound morphological disparity between the modalities: the radar RMS is a one-dimensional mechanical wave emphasizing low-frequency proximal vibrations, whereas the iPPG is an optical volume trace containing complex, high-frequency distal features (e.g., the dicrotic notch). The MS-DiCNN explicitly addresses this structural mismatch. By employing parallel dilated convolutions with varying dilation rates ($d=1, 2, 4$), the network adaptively expands its receptive field to capture both fine-grained local inflections and broad morphological envelopes without resolution loss. A Squeeze-and-Excitation (SE) attention mechanism dynamically recalibrates the channel-wise responses, forcing the network to selectively focus on the physiologically relevant morphological features while suppressing uncorrelated cross-modal noise.

\subsubsection{Modality Fusion and Subject Adaptation Module (SAM)}

Following the parallel extraction, the temporal and morphological latent vectors are integrated via an Adaptive Fusion module. A critical physiological challenge in generalized BP estimation is inter-subject variability; the parameters $a$ and $b$ in the M-K equation vary significantly across individuals due to differing arterial stiffness and baseline vascular tones.

To address this, we employ a Subject Adaptation Module (SAM). For unseen subjects during the evaluation phase, the system mimics clinical cuffless BP protocols by utilizing a brief initial calibration phase. Specifically, the chronological first 10\% of a new subject's recording is used to fine-tune the SAM, aligning the network with the individual's baseline physiological state. Following this brief adaptation, the network parameters are locked, and continuous beat-to-beat SBP and DBP estimations are performed on the remaining 90\% of the unseen data.

\section{Experimental Setup and Protocol}

\subsection{Data Collection}

As a feasibility study, the experimental validation was conducted in a controlled laboratory setting with 15 healthy volunteers (9 males, 6 females; age: $22 \pm 3$ years) with no history of genetic or cardiovascular diseases. The study was conducted in strict accordance with the Declaration of Helsinki and approved by the University of Electronic Science and Technology of China. Informed written consent was obtained from all subjects prior to data collection.

To systematically acquire the dual-modality non-contact signals alongside the continuous BP ground truth, a strictly synchronized, three-channel data acquisition framework was established (as illustrated in Figure \ref{fig1:expdesign}). The equipment and spatial configuration are detailed as follows:

\subsubsection{Distal iPPG Acquisition}
A high-precision RGB camera was positioned 50 cm directly in front of the subject's face, capturing facial video sequences at 30 Hz to monitor optical volumetric fluctuations.

\subsubsection{Proximal RMS Acquisition}
A 24 GHz FMCW mmWave radar (provided by The University of Electro-Communications, Tokyo \cite{20hoang2020noncontact}) was strategically placed 50 cm behind the subject, directly facing the posterior torso. Crucially, rather than utilizing conventional anterior chest-facing measurements, this posterior placement leverages the physical contact between the subject's back and the seat to provide a rigid boundary condition. This configuration suppresses macroscopic body sway artifacts, preserving high spectral fidelity for the subtle cardiomechanical vibrations and enabling robust, unobtrusive monitoring.

\subsubsection{Reference BP Acquisition}
A continuous non-invasive arterial pressure (CNAP) bio-signal analyzer served as the clinical ground truth. The subject's left index and middle fingers were fitted into sensor sleeves, while a calibration cuff was wrapped around the right arm to ensure precise, beat-to-beat BP calibration.

All three data streams were digitized and synchronized utilizing a BIOPAC systems platform equipped with AcqKnowledge 5.0. The intermediate frequency (IF) radar signal and the CNAP waveforms were sampled at 1000 Hz, while the iPPG signal was concurrently captured at 30 Hz.

\begin{table}[t]
\centering
\caption{EXPERIMENTAL SCHEMES FOR EACH SUBJECT}
\resizebox{\textwidth}{!}{%
\begin{tabular}{c c c}
\toprule
\textbf{State} & \textbf{Explanation} & \textbf{Duration} \\ 
\midrule
Sitting (Sit) & Normal breathing & 5 minutes \\
Deep Breathing (DB) & Breathing every 5 seconds & 5 minutes \\
Post-Exercise (Ex) & After running at 9 km/h for 2 minutes & 5 minutes \\
\bottomrule
\end{tabular}%
}
\label{tab:threeconditions}
\end{table}

To comprehensively evaluate the model's tracking capability under dynamic cardiovascular conditions, each participant was subjected to three distinct physiological perturbation protocols. Each state was continuously recorded for a 5-minute duration, as detailed in Table \ref{tab:threeconditions}. Specifically, the Sitting (Sit) protocol captured the baseline hemodynamic state with normal, spontaneous respiration. The Deep Breathing (DB) protocol intentionally induced significant Respiratory Sinus Arrhythmia (RSA) and intense intrathoracic pressure variations, challenging the model's robustness against low-frequency amplitude modulations. Finally, the Post-Exercise (Ex) protocol captured high-frequency cardiac peaks (shortened inter-beat intervals) and disordered respiratory rhythms during autonomic nervous system recovery.

\subsection{Robustness Evaluation Protocol}

In unconstrained real-world environments, distal iPPG signals are highly susceptible to ambient illumination fluctuations, sensor noises, and transmission artifacts. To rigorously evaluate the compensatory effect of the proximal RMS channel and assess the resilience of the proposed dual-modality framework, a comprehensive robustness evaluation protocol was designed, comprising three distinct evaluation phases:

\subsubsection{Comprehensive Environmental and Illumination Degradations} 
To evaluate resilience against physical environmental changes, the following synthetic degradations were systematically applied to the raw facial video streams across all three physiological states (Sit, DB, and Ex):
\begin{itemize}
    \item Environmental Noise (Gaussian): Zero-mean white Gaussian noise with a standard deviation of 0.01 was injected into each video frame, emulating stochastic background interference and sensor thermal noise.
    \item Illumination Distortions (Over/Underexposure): To simulate extreme and variable ambient lighting, frame brightness was artificially amplified (overexposure) to mimic strong direct illumination, and drastically reduced (underexposure) to emulate dim, challenging lighting environments.
\end{itemize}

\subsubsection{Targeted Telemedicine Bandwidth Artifacts}
Furthermore, to address realistic deployment challenges in remote healthcare, a targeted supplementary evaluation was conducted:
\begin{itemize}
    \item H.264 Video Compression: Standard H.264 video compression algorithms were applied specifically to the baseline resting (Sit) dataset. This specific degradation evaluates the system's robustness when macroscopic optical details are heavily compromised by low-bitrate compression artifacts in bandwidth-constrained telemedicine scenarios.
\end{itemize}

\subsubsection{Radar-Side Geometric Perturbation Sensitivity} 

To address the sensitivity of the posterior radar channel to sensing geometry, we additionally performed a post-hoc RMS perturbation analysis. Mild distance variation was approximated by RMS amplitude attenuation and SNR reduction, axis misalignment was approximated by additive phase jitter and SNR reduction, and posture-induced body sway was approximated by low-frequency baseline drift. Because the original acquisition was not designed as a controlled physical geometry-perturbation experiment, this analysis is intended as an initial sensitivity estimate rather than a substitute for future controlled experiments with actual radar-distance, angular-offset, or posture changes. 


Across all aforementioned degradation scenarios, continuous BP estimation was performed using both the dual-channel fusion network (iPPG + RMS) and an iPPG-only baseline. This comparative design explicitly quantifies the indispensable role of the mmWave radar modality in preserving prediction accuracy when the optical modality fails due to either environmental or technical constraints.
\subsection{Dataset Partitioning and Training Strategy}

To rigorously assess the generalization capability of the model and absolutely preclude data leakage, a Leave-One-Subject-Out (LOSO) cross-validation protocol was enforced across the 15 healthy participants. In this iterative scheme, the dataset was partitioned into 15 distinct folds; in each fold, all data from a single unique subject was held out exclusively for testing, while the remaining 14 subjects were utilized for model development. Specifically, within the development set of each fold, 13 subjects were assigned for training and 1 subject was dedicated to validation (for hyperparameter tuning and early stopping). To accommodate the Subject Adaptation Module (SAM) while preventing data leakage, the continuous data of the held-out test subject was chronologically split. The first 10\% of the test subject's sequence served as a personalized calibration set to initialize the SAM. The remaining 90\% of the sequence served as the true test set. Crucially, no sliding window overlap was permitted across the boundary between the 10\% calibration set and the 90\% test set, ensuring strict chronological separation and statistical independence during evaluation.

Continuous signal processing often employs sliding windows, which can inadvertently introduce statistical dependence if not handled correctly. To prevent the data leakage, the 4-second input windows (120 time steps at 30 Hz) were managed with strict isolation boundaries. While a sliding window overlap of 75\% was utilized exclusively within the training set for data augmentation, no overlapping windows or shared subject physiological features ever traversed the boundaries between the training, validation, and testing sets. The validation and testing sets utilized non-overlapping sequential windows. 

Furthermore, all normalization parameters and preprocessing statistics were calculated exclusively utilizing the training subset and subsequently applied to the held-out validation and testing subjects. Crucially, the synthetic degradations detailed in Section III.B were injected strictly after the subject-level dataset splitting. By dedicating a fully isolated validation set for hyperparameter tuning and a completely unseen testing set for final evaluation, this protocol guarantees that the reported metrics reflect the true, conservative generalization capability of the BiLSTM-MS-DiCNN model across novel populations.

\subsection{Evaluation Metrics}

To comprehensively evaluate the performance of the proposed model, we report six statistical metrics that are commonly used in cuffless BP literature and consistent with the error-reporting conventions referenced in AAMI-related guidance \cite{prsan1995aami} and IEEE Std 1708-2014 \cite{IEEE1708-2014}.


\textbf{Mean Error (ME)} represents the average signed error between predicted and true values, indicating systematic bias:

\begin{equation}
\mathrm{ME} \;=\; \frac{1}{n} \sum_{i=1}^{n} \bigl(y_i - \hat{y}_i\bigr)
\end{equation}

\textbf{Standard Deviation of Error (SDE)} quantifies the dispersion of errors around their mean, evaluating prediction stability:

\begin{equation}
\mathrm{SDE} \;=\;
\left(
\frac{1}{n-1}
\sum_{i=1}^{n}
\bigl(
(y_i - \hat{y}_i)
-
\overline{y - \hat{y}}
\bigr)^2
\right)^{\tfrac{1}{2}}
\end{equation}

\textbf{Mean Absolute Difference (MAD)} quantifies the average magnitude of absolute errors:

\begin{equation}
\mathrm{MAD} \;=\; \frac{1}{n} \sum_{i=1}^{n} \bigl\lvert y_i - \hat{y}_i \bigr\rvert
\end{equation}

In addition, we report the Root Mean Square Difference (RMSD), Mean Absolute Percentage Difference (MAPD), and Coefficient of Determination ($R^2$) as supplementary metrics. We report \textbf{ME $\pm$ SDE} and interpret them using AAMI-style reference limits (\textbf{5 $\pm$ 8 mmHg}) \cite{prsan1995aami}. MAD is also reported per IEEE Std 1708-2014 \cite{IEEE1708-2014}, where lower values (e.g., $<$ 5~mmHg) indicate stronger performance. All statistical tests are performed at the subject level (N=15) by aggregating errors per subject; sliding-window samples are not treated as independent observations.

\begin{table}[t]
\centering
\caption{Refactored performance table based on statistical indicator groups (SBP and DBP)}
\label{tab:modalityversus}
\resizebox{\textwidth}{!}{
\begin{tabular}{c c | c c c c c | c c c c c}
\toprule
\multicolumn{2}{c|}{\textbf{Dataset}} & \multicolumn{5}{c|}{\textbf{SBP Metrics (mmHg)}} & \multicolumn{5}{c}{\textbf{DBP Metrics (mmHg)}} \\
\cmidrule{1-2} \cmidrule{3-7} \cmidrule{8-12}
\textbf{Condition} & \textbf{Modality} & \textbf{ME $\pm$ SDE} & \textbf{MAD} & \textbf{RMSD} & \textbf{MAPD (\%)} & \textbf{R\textsuperscript{2}} & \textbf{ME $\pm$ SDE} & \textbf{MAD} & \textbf{RMSD} & \textbf{MAPD (\%)} & \textbf{R\textsuperscript{2}} \\
\midrule
\multirow{3}{*}{Sit} & Dual-modality & 0.56 $\pm$ 6.11 & 4.71 & 6.13 & 4.33 & 0.85 & 0.25 $\pm$ 5.76 & 4.60 & 5.76 & 4.25 & 0.87 \\
                     & iPPG-only     & -0.23 $\pm$ 8.39 & 6.29 & 8.39 & 5.60 & 0.75 & 0.33 $\pm$ 7.71 & 5.81 & 7.72 & 10.25 & 0.70 \\
                     & RMS-only      & 0.14 $\pm$ 9.73 & 7.83 & 9.74 & 7.11 & 0.62 & 0.80 $\pm$ 8.32 & 6.35 & 8.36 & 11.01 & 0.60 \\
\midrule
\multirow{3}{*}{DB}  & Dual-modality & 0.52 $\pm$ 8.20 & 6.35 & 8.22 & 5.73 & 0.76 & 0.38 $\pm$ 6.59 & 4.95 & 6.60 & 7.23 & 0.82 \\
                     & iPPG-only     & -0.12 $\pm$ 9.83 & 8.00 & 9.83 & 7.36 & 0.64 & -0.03 $\pm$ 7.84 & 6.19 & 7.84 & 9.43 & 0.75 \\
                     & RMS-only      & -0.13 $\pm$ 9.43 & 7.43 & 9.43 & 6.56 & 0.68 & 0.00 $\pm$ 8.10 & 5.99 & 8.10 & 8.92 & 0.73 \\
\midrule
\multirow{3}{*}{Ex}  & Dual-modality & -0.67 $\pm$ 7.28 & 5.33 & 7.31 & 4.40 & 0.75 & -0.42 $\pm$ 6.53 & 4.96 & 6.55 & 7.18 & 0.67 \\
                     & iPPG-only     & -0.48 $\pm$ 7.55 & 5.85 & 7.57 & 4.93 & 0.70 & 0.05 $\pm$ 6.41 & 5.11 & 6.41 & 7.37 & 0.56 \\
                     & RMS-only      & -1.38 $\pm$ 8.67 & 6.60 & 8.78 & 5.41 & 0.63 & 0.25 $\pm$ 8.18 & 6.16 & 8.18 & 9.33 & 0.46 \\
\bottomrule
\end{tabular}
}
\end{table}

\begin{figure}[htbp]
    \centering
    \begin{minipage}[b]{0.48\linewidth}
        \centering
        \includegraphics[width=1\linewidth]{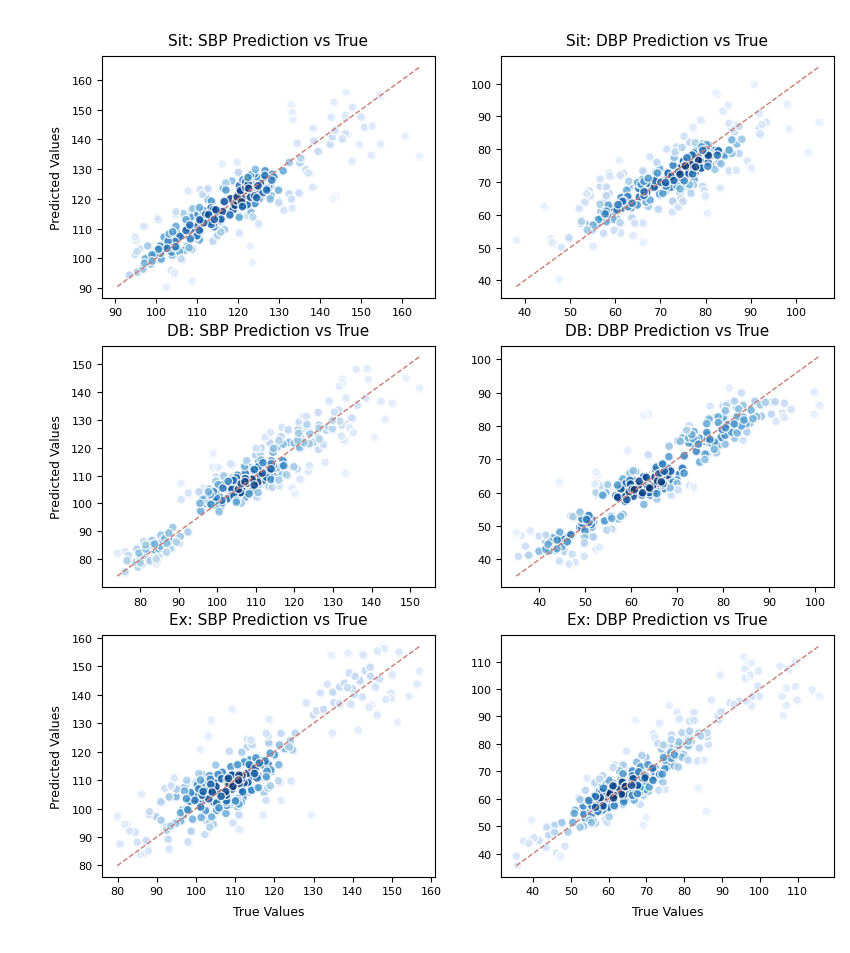}
        \caption{The scatter plot of the BP prediction value versus true value under three states.}
        \label{fig:scatter}
    \end{minipage}
    \hspace{0.02\linewidth}
    \begin{minipage}[b]{0.48\linewidth}
        \centering
        \includegraphics[width=1\linewidth]{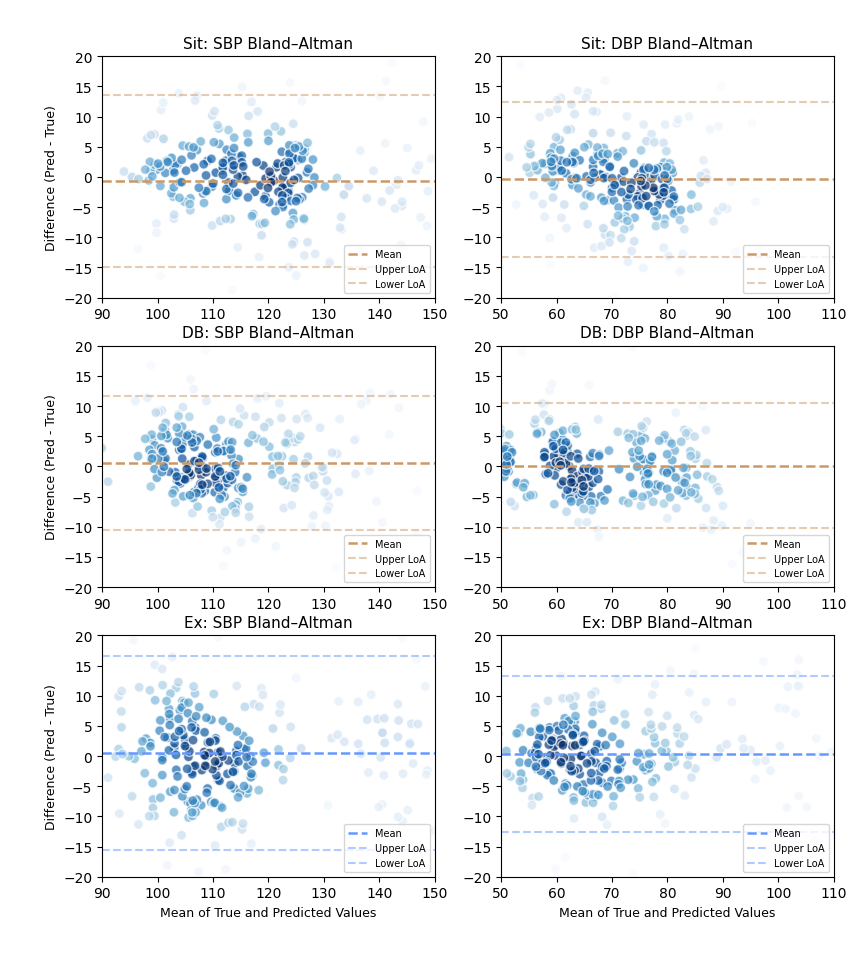}
        \caption{The Bland-Altman plot of BP estimation under three states.}
        \label{fig:Bland-Altman}
    \end{minipage}
    \vspace{-0.5em}
\end{figure}

\begin{table}
\centering
\caption{Refactored Performance Comparison of Different Models with Aggregated Metrics}
\label{table2:ippgnoised}
\resizebox{\textwidth}{!}{ 
\begin{tabular}{c c c | c c c c c | c c c c c}
\toprule
\multicolumn{3}{c|}{\textbf{Dataset}} & \multicolumn{5}{c|}{\textbf{SBP Metrics (mmHg)}} & \multicolumn{5}{c}{\textbf{DBP Metrics (mmHg)}} \\
\cmidrule{1-3} \cmidrule{4-8} \cmidrule{9-13}
\textbf{iPPG Type} & \textbf{Condition} & \textbf{Modality} & \textbf{ME $\pm$ SDE} & \textbf{MAD} & \textbf{RMSD} & \textbf{MAPD (\%)} & \textbf{R\textsuperscript{2}} & \textbf{ME $\pm$ SDE} & \textbf{MAD} & \textbf{RMSD} & \textbf{MAPD (\%)} & \textbf{R\textsuperscript{2}} \\
\midrule
 & \multirow{2}{*}{Sit} & Dual-modality  & -0.07 $\pm$ 6.90 & 5.08 & 6.90 & 4.55 & 0.83 & 0.17 $\pm$ 6.00 & 4.41 & 6.00 & 6.91 & 0.80 \\
 &                      & iPPG-only      & -2.10 $\pm$ 11.72 & 9.22 & 11.91 & 8.29 & 0.48 & -1.29 $\pm$ 9.94 & 7.64 & 9.94 & 12.59 & 0.44 \\
\cmidrule{2-13}
\multirow{2}{*}{Gauss-noise} & \multirow{2}{*}{DB} & Dual-modality  & 0.86 $\pm$ 7.82 & 6.22 & 7.87 & 5.61 & 0.78 & 0.93 $\pm$ 6.33 & 4.92 & 6.40 & 7.29 & 0.83 \\
 &                            & iPPG-only      & 0.36 $\pm$ 11.97 & 9.27 & 11.97 & 8.18 & 0.49 & -0.26 $\pm$ 11.55 & 8.67 & 11.56 & 12.90 & 0.44 \\
\cmidrule{2-13}
 & \multirow{2}{*}{Ex} & Dual-modality  & -0.06 $\pm$ 7.02 & 5.31 & 7.02 & 4.42 & 0.76 & -0.26 $\pm$ 6.35 & 4.94 & 6.36 & 7.14 & 0.67 \\
 &                     & iPPG-only      & -2.21 $\pm$ 11.05 & 8.42 & 11.27 & 6.89 & 0.39 & -1.25 $\pm$ 9.18 & 7.50 & 9.27 & 10.98 & 0.30 \\
\midrule
 & \multirow{2}{*}{Sit} & Dual-modality  & -0.17 $\pm$ 6.78 & 5.10 & 6.78 & 4.60 & 0.83 & 0.21 $\pm$ 5.82 & 4.34 & 5.83 & 6.81 & 0.81 \\
 &                      & iPPG-only      & 0.67 $\pm$ 11.53 & 8.71 & 11.55 & 7.56 & 0.55 & 0.68 $\pm$ 10.43 & 8.10 & 10.45 & 10.32 & 0.42 \\
\cmidrule{2-13}
\multirow{2}{*}{Overexposure} & \multirow{2}{*}{DB} & Dual-modality  & 0.58 $\pm$ 8.61 & 6.73 & 8.63 & 6.06 & 0.73 & 0.50 $\pm$ 6.58 & 5.01 & 6.60 & 7.27 & 0.82 \\
 &                             & iPPG-only      & 0.41 $\pm$ 11.92 & 9.28 & 11.93 & 8.20 & 0.49 & -0.23 $\pm$ 11.47 & 8.64 & 11.48 & 12.85 & 0.45 \\
\cmidrule{2-13}
 & \multirow{2}{*}{Ex} & Dual-modality  & -1.79 $\pm$ 6.73 & 5.44 & 6.96 & 4.51 & 0.73 & -2.06 $\pm$ 6.45 & 5.29 & 6.77 & 7.29 & 0.66 \\
 &                     & iPPG-only      & -2.18 $\pm$ 10.71 & 8.58 & 10.93 & 7.01 & 0.39 & -1.25 $\pm$ 8.97 & 7.81 & 9.06 & 10.12 & 0.33 \\
\midrule
 & \multirow{2}{*}{Sit} & Dual-modality  & -0.08 $\pm$ 6.98 & 5.35 & 6.98 & 4.80 & 0.82 & 0.05 $\pm$ 5.85 & 4.54 & 5.85 & 7.07 & 0.81 \\
 &                      & iPPG-only      & -1.27 $\pm$ 13.03 & 9.98 & 13.09 & 9.06 & 0.37 & -0.15 $\pm$ 11.15 & 8.64 & 11.15 & 14.45 & 0.30 \\
\cmidrule{2-13}
\multirow{2}{*}{Underexposure} & \multirow{2}{*}{DB} & Dual-modality  & 0.42 $\pm$ 7.10 & 5.67 & 7.11 & 5.12 & 0.82 & 0.25 $\pm$ 6.21 & 4.78 & 6.21 & 7.12 & 0.84 \\
 &                             & iPPG-only      & 0.55 $\pm$ 12.06 & 9.44 & 12.08 & 8.34 & 0.48 & -0.08 $\pm$ 11.63 & 8.72 & 11.63 & 13.00 & 0.44 \\
\cmidrule{2-13}
 & \multirow{2}{*}{Ex} & Dual-modality  & -0.51 $\pm$ 7.30 & 5.54 & 7.32 & 4.61 & 0.74 & -0.25 $\pm$ 6.17 & 4.82 & 6.17 & 6.97 & 0.69 \\
 &                     & iPPG-only      & -2.22 $\pm$ 11.09 & 8.49 & 11.31 & 6.94 & 0.39 & -1.19 $\pm$ 9.06 & 7.39 & 9.14 & 10.83 & 0.32 \\
\bottomrule
\end{tabular}}
\end{table}

\begin{figure}[t]
    \centering
    \includegraphics[width=1\linewidth]{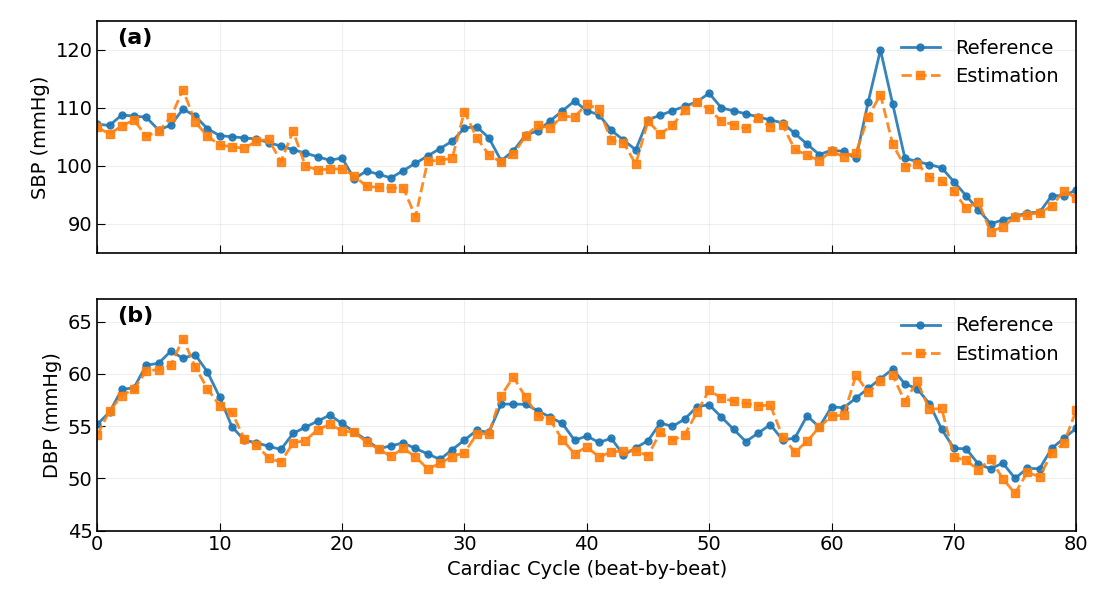}
    \caption{The comparison plot of beat-to-beat (a) SBP and (b) DBP between the proposed method and reference}
    \label{fig6: beatbybeat}
    \vspace{-1em}
\end{figure}


\section{Results}

\subsection{Dual Modality Approach Surpasses the Single Modality-based BP Estimation}

\subsubsection{Comparison Experiment under Standard Signal Quality}

Before presenting the estimation results, we first characterize the hemodynamic diversity of the dataset. Across all 15 subjects and three physiological states, the per-subject mean SBP ranges from 81.27 to 146.57~mmHg and DBP ranges from 42.17 to 103.33~mmHg. The post-exercise protocol elicits the highest mean SBP ($119.26 \pm 12.51$~mmHg), while the deep breathing protocol produces the widest DBP variability (42.17--103.33~mmHg). Notably, Subject~6 exhibits an SBP of 146.57~mmHg under the deep breathing condition, approaching the clinical hypertension threshold of 140~mmHg. This hemodynamic diversity, spanning normotensive to pre-hypertensive levels, provides a meaningful testbed for evaluating BP estimation across a physiologically relevant range.

We initially evaluated the performance of the proposed dual-modality BP estimation framework, based on the BiLSTM-MS-DiCNN architecture, under three distinct physiological conditions: Sit, DB and Ex. To provide an intuitive understanding of the results, scatter plots (Figure~\ref{fig:scatter}) and Bland-Altman plots (Figure~\ref{fig:Bland-Altman}) were used to visualize the estimation performance across these conditions. Table~\ref{tab:modalityversus} presents a detailed comparison of multiple evaluation metrics under each condition.

As illustrated in Figure~\ref{fig:scatter}, both SBP and DBP predictions closely align with the ground truth under all scenarios. Furthermore, the Bland-Altman plots of Figure \ref{fig:Bland-Altman} confirm that the majority of prediction errors lie within acceptable limits, demonstrating the robustness of our method across diverse signal dynamics. These visualizations collectively underscore the reliability and consistency of our model in both standard (Sit) and challenging (DB, Ex) conditions.
To substantiate the effectiveness of the dual-modality strategy, we conducted comparative experiments against single-modality models (iPPG-only and RMS-only).

According to Table \ref{tab:modalityversus}, under resting conditions, the dual-modality model achieved mean absolute deviations (MAD) of 4.71 mmHg (SBP) and 4.60 mmHg (DBP), both below the commonly referenced 5 mmHg MAD level used in prior cuffless BP studies. Even under more challenging scenarios---such as deep breathing and post-exercise---the model consistently maintained strong performance, with most MAD values below 6 mmHg. The only exception was the SBP MAD of 6.35 mmHg under the deep-breathing condition. This slightly higher error compared to the post-exercise state is attributed to significant intrathoracic pressure variations and chest wall displacements during deep respiration, which introduce complex low-frequency signal modulations.

Across all three datasets, the dual-modality model consistently outperformed both single-modality baselines. This superiority highlights the complementary nature of iPPG and RMS signals: while iPPG contributes rich photoplethysmographic information, RMS adds valuable time-domain mechanical features. It is also notable that the iPPG-only model generally outperformed the RMS-only model, suggesting that image-derived signals may inherently offer more comprehensive cardiovascular information than single-dimensional time-series data, particularly under standard conditions. However, as observed in Table \ref{tab:modalityversus}, the single-modality baselines (especially RMS-only) exhibit markedly higher Standard Deviations of Error (SDE $> 9$ mmHg) despite low Mean Errors. This high variance reflects the instability of unimodal prediction when tracking dynamic BP changes, further justifying the necessity of dual-modality fusion to suppress artifacts and ensure robust precision.

Additionally, as illustrated in Figure \ref{fig6: beatbybeat}, a representative beat-to-beat comparison of sit condition between the reference and estimated SBP and DBP is presented. The reference measurements yielded a mean SBP of $103.80 \pm 5.76\;\mathrm{mmHg}$ and a mean DBP of $55.24 \pm 2.75\;\mathrm{mmHg},$ whereas the proposed model produced corresponding estimates of $102.47 \pm 5.76\;\mathrm{mmHg}$ for SBP and $54.97 \pm 3.07\;\mathrm{mmHg}$ for DBP. Notably, even under conditions of rapid oscillation in the reference signal, the model reliably tracked the dynamic blood pressure fluctuations, demonstrating both high fidelity and robustness in beat-to-beat estimation.

\subsubsection{Comparison Experiment under Compromised iPPG Quality}

To further evaluate the contribution of the RMS modality in compensating iPPG signals, we designed a series of experiment where iPPG signals were corrupted with synthetic distortions, including Gaussian noise, overexposure, and underexposure. These artifacts simulate real-world challenges in non-contact monitoring, where lighting variability compromises iPPG quality. The performance of the dual-modality model under each noise condition is summarized in Table~\ref{table2:ippgnoised}, further reinforcing its resilience in low-quality signal environments.

The results in Table~\ref{table2:ippgnoised} clearly demonstrate that the proposed dual-modality framework maintains high prediction accuracy even under severe degradation of iPPG signals. Across all three types of signal distortion---additive Gaussian noise, overexposure, and underexposure---the dual-input model consistently outperforms the iPPG-only model in both SBP and DBP estimation, exhibiting minimal performance drop compared to noise-free conditions.

For instance, under Gaussian noise in the post-exercise condition, the dual-modality model achieves an SBP MAD of 5.31 mmHg and a DBP MAD of 4.94 mmHg, which are remarkably close to the baseline results without noise (5.33 mmHg and 4.96 mmHg, respectively). Similarly, even under extreme distortions (overexposure and underexposure), the model preserves low error margins and high correlation with ground truth, with $R^2$ values exceeding 0.7 in most cases.

In stark contrast, when trained and evaluated solely on the iPPG channel, the model's performance deteriorates significantly under identical distortion scenarios. For example, in the underexposure-sit condition, SBP MAD rises from 5.35 mmHg (dual-modality) to 9.94 mmHg (iPPG-only), while the $R^2$ value plummets from 0.82 to 0.37. Similar trends are observed consistently across all tested conditions and noise types, further reinforcing the vulnerability of single-modality models in compromised signal input scenarios.

These results empirically validate an important hypothesis of this study: the inclusion of mmWave radar as a complementary modality enables the model to compensate for iPPG signal degradation, thereby maintaining reliable blood pressure estimation. The dual-modality architecture not only improves overall accuracy but also significantly enhances robustness to signal noise and environmental variability. This robustness is particularly critical for the deployment of non-contact blood pressure monitoring systems in unconstrained, real-world settings, where signal quality cannot be guaranteed.

\begin{table}[t]
\centering
\caption{Statistical Metrics-Based Performance Comparison of Selected Models Under Different Conditions}
\resizebox{\textwidth}{!}{ 
\begin{tabular}{c l c c | c c c c c | c c c c c}
\toprule
\raisebox{-2pt}{\multirow{2}{*}{\textbf{DataSet}}} 
& \raisebox{-2pt}{\multirow{2}{*}{\textbf{Method}}} 
& \raisebox{-2pt}{\multirow{2}{*}{\textbf{Model Size (MB)}}} 
& \raisebox{-2pt}{\multirow{2}{*}{\textbf{Running Time (CPU / GPU)}}} 
& \multicolumn{5}{c|}{\textbf{SBP Metrics (mmHg)}} 
& \multicolumn{5}{c}{\textbf{DBP Metrics (mmHg)}} \\
\cmidrule{5-9} \cmidrule{10-14}
& & & & \textbf{ME $\pm$ SDE} & \textbf{MAD} & \textbf{RMSD} & \textbf{MAPD (\%)} & \textbf{R\textsuperscript{2}} 
  & \textbf{ME $\pm$ SDE} & \textbf{MAD} & \textbf{RMSD} & \textbf{MAPD (\%)} & \textbf{R\textsuperscript{2}} \\
\midrule
\multirow{7}{*}{Sit} & XGBoost \cite{24XGBdoi:10.15587/1729-4061.2022.265066} & 0.65 & 2.5 s / --- & \textbf{0.39} $\pm$ 14.80 & 11.47 & 14.81 & 10.89 & 0.13 & -0.33 $\pm$ 12.55 & 9.96 & 12.56 & 16.40 & 0.18 \\
& ADABoost \cite{25ADAGhosh2023BoostingAlgorithms} & 0.08 & 2.0 s / --- & -0.67 $\pm$ 14.66 & 11.48 & 14.67 & 10.93 & 0.15 & -0.48 $\pm$ 12.80 & 10.34 & 12.81 & 16.99 & 0.15 \\
& SVR \cite{EMBC2021SVR} & 1.87 & 38 s / --- & -4.31 $\pm$ 18.90 & 15.37 & 19.38 & 13.81 & -0.19 & -3.31 $\pm$ 16.42 & 13.72 & 16.75 & 22.11 & -0.41 \\
& GPR \cite{GPR_HOFD_2023} & 5.42 & 31 s / --- & -1.84 $\pm$ 17.05 & 13.41 & 17.15 & 12.27 & 0.07 & -1.65 $\pm$ 13.98 & 11.39 & 14.08 & 18.83 & 0.01 \\
& Transformer \cite{Transformer_Performer_2023} & 1.05 & 4 min / 32 s & -1.28 $\pm$ 11.33 & 8.83 & 11.40 & 8.04 & 0.59 & -0.52 $\pm$ 6.72 & 5.25 & 6.74 & 8.54 & 0.77 \\
& Dual-stream FNN & 2.08 & 7 min / 68 s & -0.76 $\pm$ 6.23 & 6.70 & 6.22 & 5.69 & 0.82 & -0.40 $\pm$ 6.01 & \textbf{4.58} & 5.78 & \textbf{4.10} & 0.85 \\
\cmidrule{2-14}
& \textbf{Our Model} & 0.80 & 4 min / 28 s & 0.56 $\pm$ \textbf{6.11} & \textbf{4.71} & \textbf{6.13} & \textbf{4.33} & \textbf{0.85} & \textbf{0.25 $\pm$ 5.76} & 4.60 & \textbf{5.76} & 4.25 & \textbf{0.87} \\
\midrule
\midrule
\multirow{7}{*}{DB} & XGBoost & 0.64 & 2.7 s / --- & -0.88 $\pm$ 17.27 & 13.91 & 17.29 & 12.38 & -0.07 & 0.14 $\pm$ 16.65 & 12.97 & 16.65 & 19.47 & -0.16 \\
& ADABoost & 0.08 & 2.8 s / --- & -0.62 $\pm$ 16.67 & 12.93 & 16.68 & 11.51 & 0.01 & -0.09 $\pm$ 15.76 & 12.31 & 15.76 & 18.55 & -0.04 \\
& SVR & 1.93 & 46 s / --- & -1.32 $\pm$ 17.96 & 13.97 & 18.01 & 12.24 & -0.16 & -0.94 $\pm$ 17.15 & 13.47 & 17.18 & 20.17 & -0.23 \\
& GPR & 6.03 & 30 s / --- & \textbf{0.19} $\pm$ 16.66 & 12.76 & 16.66 & 11.28 & 0.01 & 0.06 $\pm$ 15.42 & 12.06 & 15.42 & 18.26 & 0.01 \\
& Transformer & 1.10 & 4 min / 33 s & -0.27 $\pm$ 10.76 & 8.51 & 10.77 & 7.55 & 0.59 & -0.35 $\pm$ 7.65 & 5.94 & 7.66 & 8.92 & 0.76 \\
& Dual-stream FNN & 2.27 & 7 min / 78 s & -0.30 $\pm$ 9.50 & 6.75 & 9.55 & 6.22 & 0.73 & -0.20 $\pm$ 7.38 & 5.30 & 7.43 & 7.54 & 0.71 \\
\cmidrule{2-14}
& \textbf{Our Model} & 0.79 & 4 min / 35 s & 0.52 $\pm$ \textbf{8.20} & \textbf{6.35} & \textbf{8.22} & \textbf{5.73} & \textbf{0.76} & \textbf{0.38 $\pm$ 6.59} & \textbf{4.95} & \textbf{6.60} & \textbf{7.23} & \textbf{0.82} \\
\midrule
\midrule
\multirow{7}{*}{Ex} & XGBoost & 0.69 & 1.7 s / --- & -1.66 $\pm$ 14.28 & 11.18 & 14.38 & 9.23 & 0.01 & -0.73 $\pm$ 11.03 & 9.03 & 11.06 & 13.39 & 0.07 \\
& ADABoost & 0.10 & 1.8 s / --- & -2.26 $\pm$ 14.07 & 11.12 & 14.25 & 9.14 & 0.03 & -1.59 $\pm$ 10.72 & 8.76 & 10.84 & 12.79 & 0.10 \\
& SVR & 1.66 & 33 s / --- & -1.82 $\pm$ 15.46 & 11.86 & 15.57 & 9.78 & -0.16 & -1.57 $\pm$ 12.65 & 10.42 & 12.75 & 15.07 & -0.24 \\
& GPR & 5.50 & 27 s / --- & -1.08 $\pm$ 14.14 & 10.96 & 14.18 & 9.10 & 0.04 & -1.08 $\pm$ 11.19 & 9.24 & 11.25 & 13.54 & 0.04 \\
& Transformer & 1.04 & 4 min / 28 s & -1.16 $\pm$ 8.76 & 7.11 & 8.84 & 5.86 & 0.63 & -0.79 $\pm$ 7.07 & 5.69 & 7.12 & 8.24 & 0.61 \\
& Dual-stream FNN & 2.10 & 7 min / 62 s & -0.50 $\pm$ 8.41 & 6.11 & 8.41 & 5.14 & 0.73 & \textbf{-0.32} $\pm$ 6.87 & 4.68 & 7.01 & 6.93 & 0.65 \\
\cmidrule{2-14}
& \textbf{Our Model} & 0.88 & 3 min / 27 s & \textbf{-0.67 $\pm$} \textbf{7.28} & \textbf{5.33} & \textbf{7.31} & \textbf{4.40} & \textbf{0.75} & -0.42 $\pm$ \textbf{6.53} & \textbf{4.96} & \textbf{6.55} & \textbf{7.18} & \textbf{0.67} \\
\bottomrule
\end{tabular}}
\begin{flushleft}
\footnotesize
\end{flushleft}
\label{tab:all_model_comparison}
{\footnotesize 
\begin{flushleft}
Note: The rows with gray background indicate our model, and the bold values represent the best performance across all methods.  
Dual-stream FNN refers to the \textit{dual-stream fusion neural network} \cite{35DUALSTREAM}. 
The running time reported in this table is based on both CPU and GPU (RTX 3060).  
*Paired two-tailed tests are conducted at the subject level (N = 15) by aggregating errors per subject under each condition; improvements of our model over each baseline are statistically significant ($p < 0.05$). Sliding-window samples are not treated as independent observations.*

\end{flushleft}
\vspace{-1em}
}
\end{table}

\vspace{-0.5em}
\subsection{BiLSTM-MS-DiCNN Has the Best Performance for Accurate NCBP Based on Dual Modality}

{
To further validate the effectiveness of the proposed BiLSTM-MS-DiCNN architecture, we conducted comparative experiments against mainstream deep learning and regression baselines for non-contact, dual-modality BP estimation. Detailed metrics are summarized in Table~\ref{tab:all_model_comparison}.

Traditional regression methods such as linear regression~\cite{21tseng2020noncontact} and random forest~\cite{20hoang2020noncontact} have shown limited performance in this setting. Accordingly, we benchmarked our model against representative regressors---Extreme Gradient Boosting (XGBoost)~\cite{24XGBdoi:10.15587/1729-4061.2022.265066}, Adaptive Boosting (AdaBoost)~\cite{25ADAGhosh2023BoostingAlgorithms}, Gaussian Process Regression (GPR) \cite{GPR_HOFD_2023}, Support Vector Regression (SVR)~\cite{GPR_HOFD_2023}, and a Transformer~\cite{Transformer_Performer_2023}---as well as a dual-stream fusion neural network (FNN). 

As shown in Table~\ref{tab:all_model_comparison}, BiLSTM-MS-DiCNN delivers the best overall performance for both SBP and DBP in most scenarios. Under the resting condition, it attains the lowest SBP error (MAD = 4.71 mmHg; RMSD = 6.13 mmHg) and the highest coefficient of determination ($R^2$ = 0.85). A similar advantage holds for DBP (MAD = 4.60 mmHg; $R^2$ = 0.87). Notably, the dual-stream FNN achieves comparably strong DBP accuracy, slightly surpassing our model on DBP MAD (4.58 mmHg) and MAPD (4.10\%). Under deep-breathing, our model achieves an SBP MAD of 6.35 mmHg and a DBP MAD of 4.95 mmHg, with both $R^2$ values exceeding 0.75. Following exercise, it again outperforms all baselines, reaching an SBP MAD of 5.33 mmHg and a DBP MAD of 4.96 mmHg. These results demonstrate that our approach maintains robust accuracy across challenging conditions, outperforming the dual-stream FNN and conventional baselines in most cases.

We further compare computational efficiency by reporting model size and running time. Our model remains within 1 MB and completes inference within ~30 s on an RTX 3060 GPU for a 5-min record, offering a substantial efficiency advantage over the dual-stream FNN.

Overall, BiLSTM-MS-DiCNN achieves a favorable trade-off between accuracy and efficiency. While classical regressors (XGBoost, AdaBoost, SVR, GPR) are lighter and faster, their BP estimation accuracy is markedly lower. The dual-stream FNN approaches our accuracy but at a higher computational cost. By jointly modeling temporal dependencies via the BiLSTM and enhancing local feature representations with the MS-DiCNN, our ``divide-and-conquer'' design effectively fuses spatial and temporal information for accurate and resilient BP estimation under real-world, unconstrained conditions, supporting continuous NCBP applications.
}
\vspace{-0.8em}

\begin{table}
\centering
\caption{Ablation Study Results of Proposed Model Under Sit Condition}
\resizebox{\textwidth}{!}{ 
\begin{tabular}{cc|ccccc|ccccc}
\toprule
\multirow{2}{*}{\textbf{Experiment ID}} & \multirow{2}{*}{\textbf{Ablation Module}} 
& \multicolumn{5}{c|}{\textbf{SBP Metrics (mmHg)}} 
& \multicolumn{5}{c}{\textbf{DBP Metrics (mmHg)}} \\
\cmidrule{3-7} \cmidrule{8-12}
& & \textbf{ME $\pm$ SDE} & \textbf{MAD} & \textbf{RMSD} & \textbf{MAPD (\%)} & \textbf{R\textsuperscript{2}}
  & \textbf{ME $\pm$ SDE} & \textbf{MAD} & \textbf{RMSD} & \textbf{MAPD (\%)} & \textbf{R\textsuperscript{2}} \\
\midrule
Experiment a  & MS-DiCNN     & -2.05 $\pm$ 16.32 & 12.54 & 16.45 & 11.22 & 0.14 & -1.36 $\pm$ 13.22 & 10.44 & 13.28 & 16.69 & 0.11 \\
Experiment b  & BiLSTM       & -0.32 $\pm$ 15.51 & 10.66 & 15.51 & 9.53  & 0.24 & -0.60 $\pm$ 11.03 & 8.59  & 11.05 & 13.99 & 0.39 \\
Experiment c  & Fusion       & -0.05 $\pm$ 8.49  & 6.37  & 8.49  & 5.87  & 0.72 & -0.21 $\pm$ 7.93  & 5.80  & 7.94  & 9.73  & 0.69 \\
Experiment d  & Dilation     & -2.60 $\pm$ 16.39 & 12.86 & 16.59 & 11.64 & 0.13 & -1.79 $\pm$ 13.52 & 11.18 & 13.63 & 17.96 & 0.07 \\
Experiment e  & LN+Dropout   & -1.30 $\pm$ 20.96 & 16.07 & 21.00 & 15.30 & -0.40 & -1.64 $\pm$ 17.31 & 15.46 & 17.39 & 21.55 & -0.52 \\
\midrule
\textbf{Our Model} & \textbf{None} & \textbf{0.56 $\pm$ 6.11} & \textbf{4.71} & \textbf{6.13} & \textbf{4.33} & \textbf{0.85} & \textbf{0.25 $\pm$ 5.76} & \textbf{4.60} & \textbf{5.76} & \textbf{4.25} & \textbf{0.87} \\
\bottomrule
\end{tabular}}
\begin{flushleft}
\footnotesize
\end{flushleft}
\label{tab:ablation_study}

\vspace{-0.5mm}
{\footnotesize 
\begin{flushleft}
Note: Experiment a and b remove the MS-DiCNN branch and the BiLSTM branch of the main architecture, respectively. Experiment c substitute the fusion weights of two channels into an average superposition. Experiment d substitute the dilated CNN to a standard CNN. Experiment e removes the layer normalization and dropout layer.
\end{flushleft}
}
\vspace{-1em}
\end{table}

\subsection{Each Module of BiLSTM-MS-DiCNN is Important According to Ablation Experiment}

To systematically evaluate the contribution of each architectural component in our BiLSTM-MS-DiCNN framework, we conducted a series of ablation experiments under the sit condition. Each experiment modifies or removes a specific module---including the BiLSTM, the MS-DiCNN, the cross-modal fusion strategy, the dilation mechanism, and the regularization components (Layer Normalization (LN) + Dropout)---to quantify its individual impact on model performance.

The results of these ablation studies are summarized in Table~\ref{tab:ablation_study}. The complete model (denoted as Our Model), which integrates all six components, consistently achieves the best performance, with the lowest prediction errors for both SBP (MAD = 4.71 mmHg, RMSD = 6.13 mmHg) and DBP (MAD = 4.60 mmHg, RMSD = 5.76 mmHg), as well as the highest correlation scores ($R^2$ = 0.85 for SBP and 0.87 for DBP).

When the MS-DiCNN block is removed (Experiment a), the performance degrades drastically---SBP MAD increases to 12.54 mmHg and DBP MAD to 10.44 mmHg---demonstrating the critical role of multi-scale spatial convolutions in capturing fine-grained features relevant to blood pressure dynamics.

Eliminating the BiLSTM branch (Experiment b) results in a similarly pronounced performance drop (SBP MAD = 10.66 mmHg), confirming the necessity of modeling long-range temporal dependencies for accurate BP estimation.

In Experiment c, we replaced the proposed fusion mechanism with a simple averaging operation. Although this modification yields better results than other ablated configurations, it still underperforms compared to the complete model (SBP MAD = 6.37 mmHg, $R^2$ = 0.72), underscoring the importance of a learnable, modality-aware fusion strategy in effectively integrating complementary iPPG and RMS features.

Disabling the dilation mechanism in the convolutional layers (Experiment d) further increases SBP MAD to 12.86 mmHg, revealing that the enlarged receptive field afforded by dilated convolutions is essential for robust spatial feature extraction across multiple temporal scales, compared with a standard CNN.

Lastly, removing both Layer Normalization and Dropout (Experiment e) leads to the worst overall performance, with SBP MAD rising to 16.07 mmHg and DBP MAD to 15.46 mmHg, and both $R^2$ values turning negative. This outcome clearly highlights the necessity of proper regularization for mitigating overfitting and ensuring training stability.

In summary, the ablation results conclusively demonstrate that each module or strategy within the BiLSTM-MS-DiCNN architecture plays a vital and non-redundant role. The combination of temporal modeling, multi-scale spatial feature extraction, modality-aware fusion, and effective regularization yields a robust and generalizable framework for non-contact blood pressure estimation, particularly under real-world physiological variability.

\subsection{Robustness Analysis under Video Compression}

To evaluate the system's robustness against video compression artifacts---a critical challenge in low-bandwidth telemedicine---we conducted a video-level compression experiment on the sit dataset. We applied H.264 compression algorithm to the raw video recordings and re-extracted iPPG signals from the compressed video streams. These degraded signals were then evaluated using the same pre-trained models to compare the resilience of the standalone single-channel iPPG approach versus the proposed dual-channel Fusion framework.

Figure 7 presents the performance comparison. Under the compressed condition, the single-modality iPPG model exhibited a significant systemic bias shift. Specifically, the Mean Error (ME) for SBP drifted from -0.23 mmHg to -2.50 mmHg, indicating a tendency to underestimate blood pressure when optical details are lost. In contrast, the proposed Fusion model demonstrated superior robustness. Although a slight performance drop was observed, the Fusion model maintained a much lower systemic bias (ME = -1.09 mmHg) and consistent precision (SDE = 6.33 mmHg), validating that the mmWave radar modality effectively compensates for the degradation of optical signals.

\begin{figure}[h]
    \centering
    \includegraphics[width=1\linewidth]{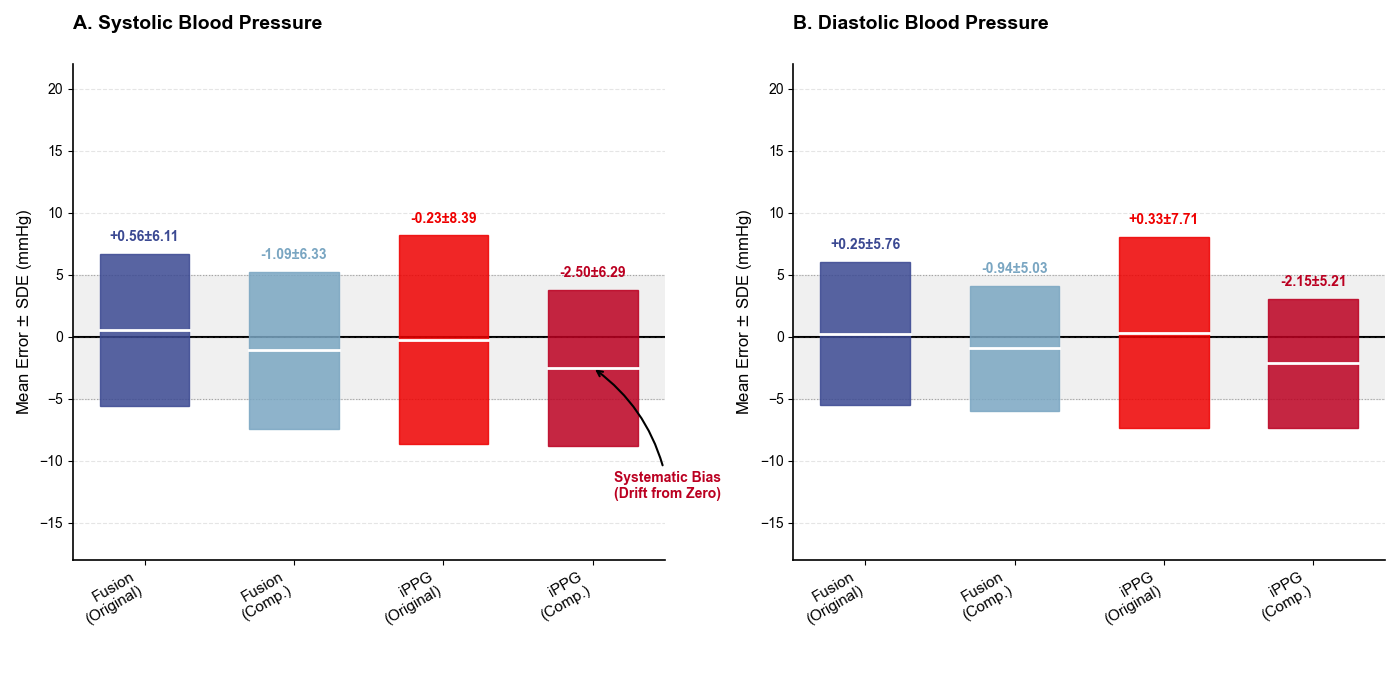}
    \caption{Statistical comparison of (a) SBP (b) DBP estimation errors under different signal qualities.}
    \label{fig:placeholder}
\end{figure}

\subsection{Calibration Sensitivity of the Subject Adaptation Module}

To further examine the operational role of the Subject Adaptation Module (SAM), we conducted a calibration-ratio sensitivity analysis by varying the amount of subject-specific initialization data from 0\% to 10\%. This experiment directly evaluates whether the proposed framework depends entirely on subject-specific calibration or whether it retains baseline generalization capability when SAM is disabled. The results are summarized in Table VI.

\begin{table}[t]
\centering
\caption{Model performance under different calibration ratios.}
\label{tab:calibration_ratio}
\setlength{\tabcolsep}{3pt} 
\resizebox{\textwidth}{!}{
\begin{tabular}{ccccccc}
\hline
\textbf{Cali.} & \textbf{SBP MAD} & \textbf{DBP MAD} & \textbf{SBP SDE} & \textbf{DBP SDE} & \textbf{SBP}   & \textbf{DBP}   \\
\textbf{Ratio}       & \textbf{(mmHg)}  & \textbf{(mmHg)}  & \textbf{(mmHg)}  & \textbf{(mmHg)}  & \textbf{R\textsuperscript{2}} & \textbf{R\textsuperscript{2}} \\
\hline
0\%         & 7.52    & 6.85    & 8.52    & 7.85    & 0.72  & 0.74  \\
2\%         & 6.45    & 5.98    & 7.68    & 7.12    & 0.76  & 0.78  \\
4\%         & 5.73    & 5.42    & 7.05    & 6.58    & 0.79  & 0.81  \\
6\%         & 5.21    & 5.05    & 6.62    & 6.18    & 0.82  & 0.84  \\
8\%         & 4.92    & 4.78    & 6.35    & 5.92    & 0.84  & 0.86  \\
10\%        & 4.71    & 4.60    & 6.11    & 5.76    & 0.85  & 0.87  \\
\hline
\end{tabular}
}
\end{table}

As shown in Table VI, the model maintains a measurable non-calibrated baseline capability even when the calibration ratio is reduced to 0\%, achieving SBP and DBP MADs of 7.52 mmHg and 6.85 mmHg, respectively. This result indicates that the backbone dual-modality architecture can still extract generalizable hemodynamic representations without subject-specific initialization. Nevertheless, calibration consistently improves both error magnitude and prediction stability. Increasing the calibration ratio from 0\% to 10\% reduces SBP MAD from 7.52 mmHg to 4.71 mmHg and DBP MAD from 6.85 mmHg to 4.60 mmHg, while also decreasing SDE and improving $R^2$ for both SBP and DBP.

The performance gain is most pronounced at low calibration ratios and gradually saturates as the calibration ratio approaches 10\%, suggesting diminishing returns beyond brief subject-specific initialization. These findings clarify that SAM should be interpreted as a calibration-based personalization mechanism rather than a prerequisite for model functionality. In other words, calibration improves subject-level adaptation and reduces inter-subject variability, but the proposed framework is not solely enabled by the calibration process. This analysis also supports the practical feasibility of the protocol, as the default 10\% calibration corresponds to only a short initial recording segment before continuous BP estimation is performed on the remaining unseen data.

\section{Discussion}

This study investigates the feasibility of a dual-modality BP estimation framework based on iPPG and RMS, implemented via the proposed BiLSTM-MS-DiCNN architecture. By combining temporal modeling (BiLSTM) with multi-scale spatial feature extraction (MS-DiCNN), the model captures complementary information across heterogeneous signals. Evaluated on a controlled dataset from 15 healthy participants across three physiological states, the framework achieved performance within commonly cited error bounds reported in prior cuffless BP studies \cite{prsan1995aami,IEEE1708-2014}, supporting the feasibility of dual-modality NCBP monitoring.

Limitations. Several limitations of this feasibility study should be acknowledged. First, the present dataset was collected from a limited cohort of 15 young, healthy volunteers under controlled laboratory settings, without large-scale clinical validation against invasive or oscillometric references. While the three physiological protocols (resting, deep breathing, and post-exercise) collectively elicit a wide BP range (SBP: 81--147~mmHg, DBP: 42--103~mmHg across per-subject means), encompassing normotensive to pre-hypertensive levels, this cohort does not include elderly subjects, hypertensive patients, or diverse ethnic populations. Second, although the proposed dual-modality framework effectively mitigates the optical vulnerabilities of iPPG, its performance may still be affected by extreme illumination conditions, motion artifacts, and inter-subject variations such as skin tone and vascular depth. Third, the current study did not explicitly examine the effect of the Signal Quality Index (SQI) threshold used for signal selection. Fourth, the CNAP finger-cuff system used as ground truth is susceptible to localized vasoconstriction and hydrostatic pressure artifacts, particularly during the post-exercise protocol when peripheral vascular tone undergoes rapid fluctuations. The manufacturer-specified MAP accuracy of the CNAP system indicates that reference-device uncertainty is non-negligible. Therefore, the reported SBP/DBP MAD values should be interpreted as model-to-CNAP discrepancies rather than absolute intra-arterial BP errors; part of the residual error may arise from reference uncertainty, especially during post-exercise recovery. 

Regarding the SAM, we emphasize that brief subject-specific calibration is a standard and widely accepted practice in cuffless BP monitoring, as recommended by IEEE Std~1708-2014 \cite{IEEE1708-2014}. To validate that the SAM enhances performance through standard domain adaptation rather than compensating for architectural deficiency, we conducted a calibration ratio analysis (0\%--10\%). Critically, even without any calibration (0\%), the model achieves SBP MAD of 7.52~mmHg and DBP MAD of 6.85~mmHg, which remain within the range reported by state-of-the-art single-modality methods.
Performance improves with diminishing returns as calibration increases, indicating that calibration improves subject-level personalization and reduces inter-subject variability, rather than serving as the sole source of model capability.

Despite these limitations, the feasibility study provides several valuable insights. To quantify the value of dual-modality input, we compared against single-modality baselines using either iPPG or RMS alone (Table II). In the Sit scenario, our iPPG-only baseline attains SBP: $-0.23 \pm 8.39$ and DBP: $0.33 \pm 7.71$, which is comparable to a single-channel iPPG state of the art \cite{0913-1}. The RMS-only baseline yields SBP: $0.14 \pm 9.73$ and DBP: $0.80 \pm 8.32$. Notably, prior single-modal mmWave work reports excellent accuracy under short-range, stationary, tightly controlled protocols---SBP: $0.87 \pm 5.01$; DBP: $1.55 \pm 5.27$ \cite{0908-7}---which are not directly comparable to our meter-scale, motion-aware setting \cite{42,44}. While direct benchmarking is challenging due to the scarcity of synchronized dual-modality (iPPG and mmWave radar) BP datasets, we contextualize our results by comparing them with recently reported non-contact BP estimation methods. Mainstream iPPG-based methods utilizing deep learning typically report SBP MAD ranging from 5.0 to 8.0 mmHg in controlled settings \cite{17rong2021blood, 0913-1, 15park2024robust}. Similarly, continuous radar-based cardiovascular monitoring studies often report MADs around 6.0 to 9.0 mmHg \cite{0908-7, 43, 35DUALSTREAM}. In this study, our dual-modality model achieved SBP/DBP MADs of 4.71/4.60 mmHg, with error dispersions (SDEs) of 6.11/5.76 mmHg. These results fall within commonly cited AAMI-style reference error bounds \cite{prsan1995aami} and are competitive with recent non-contact BP estimation methods, supporting the viability of the dual-modality approach.

We further assessed robustness under real-world degradations by injecting controlled impairments into the iPPG channel (Gaussian noise, overexposure, underexposure), which have been reported by recent research that degraded iPPG could negatively impact the BP estimation \cite{0913-2,45,47,50}. As summarized in Table III, even when iPPG quality was severely degraded, the dual-input model maintained reliable performance and clearly outperformed the iPPG-only counterpart. For example, under Gaussian-noise in the DB scenario, the iPPG-only SDE rises to around 12 mmHg, whereas the dual-modality model remains lower than 8 mmHg, highlighting RMS as a complementary cue when optical signals deteriorate \cite{47}.

Beyond input modality, we examined the architectural advantage of BiLSTM-MS-DiCNN by comparing it with several mainstream models. Across benchmarks, our model achieves competitive results across most metrics---averaging SBP MAD = 5.46 $\pm$ 0.83 mmHg and DBP MAD = 4.84 $\pm$ 0.20 mmHg---suggesting effective fusion of temporal and local temporal patterns \cite{48,50}. While a dual-stream FNN attains relatively strong accuracy, our architecture offers higher computational efficiency. This achieves a similar effect as the practice of using distributed mmWave modules for BP estimation \cite{0913-3,43}.
Ablation studies confirm that each component is necessary for peak performance: removing BiLSTM, MS-DiCNN, dilation, fusion, or regularization consistently degrades accuracy and stability. The synergy among these modules supports generalization and reliable convergence across all conditions \cite{48}.

Physiological Interpretability. A critical concern regarding end-to-end deep learning models is whether the learned representations correspond to physiologically meaningful quantities. To address this, we conducted a linear probe analysis to evaluate whether the intermediate features of BiLSTM-MS-DiCNN encode PTT information. Specifically, we extracted the fused feature vectors from the trained model and trained a Ridge regression to predict non-contact PTT (ncPTT)---computed via cross-correlation between the iPPG and RMS signals---from these features. The results demonstrate that the learned features can predict ncPTT with a Pearson correlation coefficient of $r = 0.726$ ($p < 10^{-30}$), indicating that approximately 51\% of the ncPTT variance is linearly decodable from the model's internal representation. This provides direct evidence that the BiLSTM-MS-DiCNN network implicitly captures the physiologically relevant timing relationship between proximal cardiac ejection and distal pulse arrival, which is the fundamental principle underlying PTT-based blood pressure estimation. Importantly, this implicit encoding is more robust than explicit peak-detection-based PTT extraction, as it is resilient to signal noise and morphological disparities between heterogeneous modalities.

Furthermore, to verify that the network tracks physiologically meaningful patterns rather than cross-modal noise, we analyzed the internal attention mechanisms of BiLSTM-MS-DiCNN. The AdaptiveFusion module learnable weights converge to $\alpha = 0.499$ (iPPG, 46.0\%) and $\beta = 0.587$ (RMS, 54.0\%), indicating balanced modality contribution with a slight preference for the proximal cardiac mechanical signal. Multi-scale branch activation analysis reveals modality-specific dilation preferences: the iPPG branch exhibits the highest activation at $d=4$ (2.29), capturing broad pulse wave morphology, while the RMS branch peaks at $d=2$ (0.35), focusing on intermediate-scale cardiac vibrations. Feature magnitude analysis further shows that iPPG relies predominantly on morphological features (CNN magnitude 2.08 vs. LSTM 0.67), whereas RMS relies on temporal features (LSTM 0.52 vs. CNN 0.32). These results provide converging evidence that the network learns physiologically consistent, modality-specific representations.

Future work will focus on expanding the dataset to include diverse demographics and pathological cases, conducting clinical trials with certified medical devices, and integrating adaptive calibration. These efforts will further advance non-contact, multimodal blood-pressure monitoring toward reliable clinical and wearable deployment \cite{43,49}.

\section{Conclusion}

This study investigates the feasibility of a dual-modality non-contact blood pressure (NCBP) estimation framework that integrates facial imaging photoplethysmography (iPPG) and posterior-facing millimeter-wave radar signals. The proposed BiLSTM-MS-DiCNN architecture was designed to extract both global temporal dependencies and local spatial features from these heterogeneous physiological inputs, overcoming the morphological disparity between optical and radar modalities. Experiments across sitting, deep-breathing, and post-exercise scenarios demonstrated promising performance, with the dual-modality model achieving SBP/DBP MADs of 4.71/4.60 mmHg under resting conditions, falling within commonly referenced AAMI-style error bounds. Comparative experiments against single-modality baselines confirmed the complementary nature of iPPG and radar inputs, while robustness evaluations under signal degradation (Gaussian noise, overexposure, underexposure, and video compression) demonstrated the compensatory role of the radar modality when optical signals are compromised. Ablation studies further confirmed the essential contributions of each architectural module. Importantly, a linear probe analysis revealed that the learned features encode pulse transit time information (Pearson $r = 0.726$, $p < 10^{-30}$), providing evidence that the model implicitly captures physiologically relevant timing relationships. While the current feasibility study is limited to a controlled cohort of 15 healthy young participants, the results support the viability of mmWave-iPPG dual-modality fusion as a promising pathway toward robust, unobtrusive NCBP monitoring. Future work will focus on expanding the dataset to include diverse demographics and pathological cases, conducting clinical trials with certified medical devices, and integrating adaptive calibration strategies.

\begin{figure}[t]
    \centering
    \includegraphics[width=0.85\textwidth]{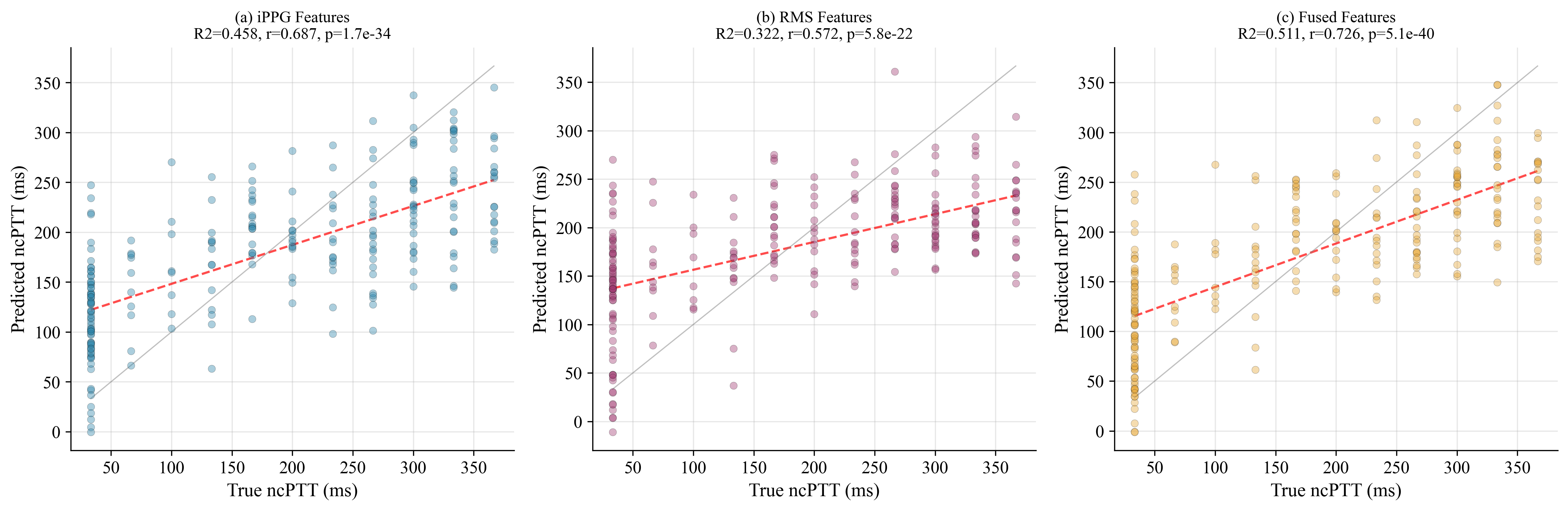}
    \caption{Linear probe results: true ncPTT vs. predicted ncPTT for (a)~iPPG features, (b)~RMS features, and (c)~fused features. The fused features achieve the highest correlation ($r=0.726$, $R^2=0.511$), confirming that the model's internal representation encodes PTT-related timing information.}
    \label{fig:linear_probe}
\end{figure}

\end{document}